%% file: DiPhotonMetPRL.tex
\documentclass[preprint,twocolumn,5p]{elsarticle}

\usepackage{graphicx}
\usepackage{xspace}
\usepackage{units}
\usepackage{textpos}
\usepackage{amssymb}
\usepackage{hyperref}

\biboptions{compress}

\newcommand{\CL}{CL\xspace}

\newcommand{\MET}{\ensuremath{E_{\mathrm{T}}^{\mathrm{miss}}}\xspace}
\newcommand{\met}{\MET}
\newcommand{\Section}{Section}
\newcommand{\Fig}{Fig.\xspace}
\newcommand{\Ref}{Ref.\xspace}
\newcommand{\Tab}{Table\xspace}

\newcommand{\mass}[1]{\ensuremath{m_{#1}}}
\newcommand{\ET}{\ensuremath{E_{\mathrm{T}}}\xspace}
\newcommand{\pt}{\ensuremath{p_{\mathrm{T}}}\xspace}

\newcommand{\antibar}[1]{\ensuremath{#1\bar{#1}}}
\newcommand{\ttbar}{\antibar{t}\xspace}
\newcommand{\Zee}{\ensuremath{Z \rightarrow ee}\xspace}
\newcommand{\myW}{\ensuremath{W(\to e\nu)+\mathrm{jets}/\gamma}\xspace}
\newcommand{\myttbar}{\ensuremath{\ttbar(\to e\nu)+\mathrm{jets}}\xspace}
\newcommand{\Zgg}{\ensuremath{Z(\to \nu\bar{\nu})+\gamma\gamma}\xspace}
\newcommand{\Wgg}{\ensuremath{W(\to \ell\nu)+\gamma\gamma}\xspace}

\newcommand{\ifb}{\mbox{fb$^{-1}$}}
\newcommand{\ipb}{\mbox{pb$^{-1}$}}

\newcommand{\Nfive}{\ensuremath{N_{5}}\xspace}

\newcommand{\QCDg}{\ensuremath{\mathrm{QCD}}\xspace}
\newcommand{\neutralino}{\ensuremath{\tilde{\chi}^{0}_{1}}\xspace}
\newcommand{\neutralinotwo}{\ensuremath{\tilde{\chi}^{0}_{2}}\xspace}
\newcommand{\chargino}{\ensuremath{\tilde{\chi}^{\pm}_{1}}\xspace}
\newcommand{\gravitino}{\ensuremath{\tilde{G}}\xspace}
\newcommand{\gluino}{\ensuremath{\tilde{g}}\xspace}
\newcommand{\Pythia}{{\tt PYTHIA}\xspace}
\newcommand{\Herwigpp}{{\tt HERWIG++}\xspace}
\newcommand{\Madgraph}{{\tt MadGraph}\xspace}
\newcommand{\Suspect}{{\tt SUSPECT}\xspace}
\newcommand{\Sdecay}{{\tt SDECAY}\xspace}
\newcommand{\Isajet}{{\tt ISAJET}\xspace}
\newcommand{\Prospino}{{\tt PROSPINO}\xspace}

\newcommand{\integLumi}{\unit[1.07]{\ifb}\xspace}
\newcommand{\integLumiE}{\unit[$(1.07\pm 0.04)$]{\ifb}\xspace}
\newcommand{\Nlimit}{\ensuremath{7.1}\xspace}

\newcommand{\UEDlimitlow}{\unit[1.23]{TeV}\xspace}

\newcommand{\GGMlimit}{\unit[805]{GeV}\xspace}

\newcommand{\SPSlimit}{\unit[145]{TeV}\xspace}

\newcommand{\Nsignal}{5\xspace}
\newcommand{\nbackground}{\ensuremath{4.1 \pm 0.6 (\mathrm{stat}) \pm 1.6 (\mathrm{syst})}\xspace}
\newcommand{\sigmaGGM}{\ensuremath{\sigma < \unit[(22-129)]{fb}}\xspace}
\newcommand{\sigmaUED}{\ensuremath{\sigma < \unit[(15-27)]{fb}}\xspace}
\newcommand{\sigmaBRUED}{\ensuremath{\sigma < \unit[(13-15)]{fb}}\xspace}
\newcommand{\sigmaSPS}{\ensuremath{\sigma < \unit[(27-91)]{fb}}\xspace}

\journal{Physics Letters B}

\begin{document}

\begin{frontmatter}

\title{Search for Diphoton Events with Large Missing Transverse Momentum
in \unit[1]{\ifb} of \unit[7]{TeV} Proton-Proton Collision Data with the ATLAS Detector
}

\author{The ATLAS Collaboration}
\address{CERN, 1211 Geneva 23, Switzerland }
\ead{atlas.publications@cern.ch}

\begin{abstract}
  \begin{textblock}{6}[0,1](-0.1,-3.5){\makebox[0pt][l]{\small CERN-PH-EP-2011-160}}\end{textblock}
  A search for diphoton events with large missing transverse momentum
  has been performed using \integLumi of proton-proton collision data at $\sqrt{s} =
  \unit[7]{TeV}$ recorded with the ATLAS detector. No excess of events was observed above the Standard Model prediction and 
  \unit[95]{\%}~Confidence Level (\CL) upper limits are set
  on the production cross section for new physics. The limits depend
  on each model parameter space and vary as follows: \sigmaGGM in the context of a generalised
  model of gauge-mediated supersymmetry breaking (GGM) with a
  bino-like lightest neutralino,  \sigmaSPS in the context of a
  minimal model of gauge-mediated supersymmetry breaking (SPS8),
  and  \sigmaUED in the context of a specific model with
  one universal extra dimension (UED). A \unit[95]{\%}~\CL lower limit of \GGMlimit, for
  bino masses above \unit[50]{GeV}, is set on the GGM gluino mass. Lower
  limits of \SPSlimit and \UEDlimitlow are set on the SPS8 breaking scale $\Lambda$ and 
  on the UED compactification scale $1/R$, respectively.
  These limits provide the most stringent tests of these models to date.
\end{abstract}

\end{frontmatter}

\section{Introduction}
This Letter reports on the search for diphoton ($\gamma\gamma$) events
with large missing transverse momentum (\MET) in \integLumi of
proton-proton ($pp$) collision data at $\sqrt{s}=\unit[7]{TeV}$
recorded with the ATLAS detector in the first half of 2011, extending
a prior study performed with \unit[36]{\ipb}~\cite{Aad:2011kz}.  The
results are interpreted in the context of three models of new physics:
a general model of gauge-mediated supersymmetry breaking
(GGM)~\cite{Meade:2008wd,Buican:2008ws,Ruderman:2011vv}, a minimal
model of gauge-mediated supersymmetry breaking
(SPS8)~\cite{Allanach:2002nj}, and a model positing one universal
extra dimension
(UED)~\cite{Appelquist:2000nn,Macesanu:2002ew,Macesanu:2005jx}.

\section{Supersymmetry}
\label{sec:susy}
Supersymmetry~(SUSY)~\cite{Golfand:1971iw,Neveu:1971rx,Ramond:1971gb,Volkov:1973ix,Wess:1974tw}
introduces a symmetry between fermions and bosons, resulting in a SUSY
partner (sparticle) with identical quantum numbers except a difference
by half a unit of spin for each Standard Model (SM) particle. As none
of these sparticles have been observed, SUSY must be a broken symmetry
if realised in nature.  Assuming $R$-parity
conservation~\cite{Fayet:1977yc,Farrar:1978xj}, sparticles have to be
produced in pairs.  These would then decay through cascades involving
other sparticles until the lightest SUSY particle (LSP) is produced,
which is stable.

In gauge-mediated SUSY breaking~(GMSB)
models~\cite{Dine:1981,Dimopoulus:1981,AlvarezGaume:1981wy,Nappi:1982,Dine:1995ag}
the LSP is the gravitino \gravitino. GMSB experimental signatures are
largely determined by the nature of the next-to-lightest SUSY
particle~(NLSP), which for a large part of the GMSB parameter space is
the lightest neutralino \neutralino. Should the lightest neutralino
have similar couplings as the SM U(1) gauge boson, also referred to as
``bino'' in this case, the final decay in the cascade would
predominantly be $\neutralino\to\gamma\gravitino$, with two cascades
per event, leading to final states with $\gamma\gamma+\met$, where
$\met$ results from the undetected gravitinos.

Searches for GMSB performed at the Tevatron~\cite{Aaltonen:2009tp,Abazov:2010us}
were optimized to be sensitive to a minimal GMSB model
(SPS8)~\cite{Allanach:2002nj}. To reduce the number of free parameters
in this model, several assumptions are made. These assumptions lead to
a mass hierarchy in which squarks and gluinos are much heavier than
the lightest neutralino and chargino \chargino. The SUSY breaking mass
scale felt by the low-energy sector, $\Lambda$, is the only free
parameter of the SPS8 model.
The other model parameters are fixed to the following values: the
messenger mass $M_{\mathrm{mess}}=2\Lambda$, the number of copies of
$5+\bar{5}$ SU(5) messengers $\Nfive=1$, the ratio of the 
vacuum expectation values of the two Higgs doublets $\tan\beta=15$, and the Higgs sector mixing
parameter $\mu>0$. The NLSP is assumed to decay promptly
($c\tau_{\mathrm{NLSP}} < \unit[0.1]{mm}$).
At the present LHC energy the main contribution to the production
cross section in the SPS8 model is via gaugino pair production, i.e.
production of \neutralinotwo\chargino or \neutralinotwo\neutralinotwo
pairs.  The contribution from gluino and/or squark pairs is below
\unit[10]{\%} of the production cross section due to their high
masses.
Besides the two photons and the two gravitinos, jets, leptons, and gauge
bosons may be produced in the cascades.
This Letter presents the first limits on the SPS8 model at the LHC.
Furthermore, a GGM SUSY model is considered in which the gluino and
neutralino masses are treated as free parameters.  The other sparticle
masses are fixed at $\sim\unit[1.5]{TeV}$, leading to a dominant
production mode at $\sqrt{s} = \unit[7]{TeV}$ of a pair of gluinos via
the strong interaction that would decay via cascades into the
bino-like neutralino NLSP. Jets may be produced in the cascades from
the gluino decays if kinematically allowed. Further model parameters
are fixed to $\tan\beta=2$ and $c\tau_{\mathrm{NLSP}} <
\unit[0.1]{mm}$.
The decay into the wino-like neutralino NLSP is possible and was
studied by the CMS Collaboration~\cite{Chatrchyan:2011ah}.

\section{Extra dimensions}
UED models postulate the existence of additional spatial dimensions in
which all SM particles can propagate, leading to the existence of a
series of excitations for each SM particle, known as a Kaluza-Klein
(KK) tower. This analysis considers the case of a single UED, with
compactification radius (size of the extra dimension) $R \approx
\unit[1]{TeV^{-1}}$.
At the LHC, the main UED process would be the production via the
strong interaction of a pair of first-level KK quarks and/or
gluons~\cite{Macesanu:2002db}. These would decay via cascades
involving other KK particles until reaching the lightest KK particle
(LKP), i.e. the first level KK photon $\gamma^*$. SM particles as
quarks, gluons, leptons, and gauge bosons may be produced in the
cascades. If the UED model is embedded in a larger space with $N$
additional eV$^{-1}$-sized dimensions accessible only to
gravity~\cite{DeRujula:2000he}, with a $(4 + N)$-dimensional Planck
scale ($M_D$) of a few \unit{TeV}, the LKP would decay gravitationally
via $\gamma ^* \rightarrow \gamma + G$. $G$ represents a tower of
eV-spaced graviton states, leading to a graviton mass between $0$ and
$1/R$.
With two decay chains per event, the final state would contain $\gamma
\gamma + \MET$, where \MET results from the escaping gravitons.
Up to $1/R \sim\unit[1]{TeV}$, the branching ratio to the diphoton and
\MET final state is close to \unit[100]{\%}. As $1/R$ increases, the
gravitational decay widths become more important for all KK particles
and the branching ratio into photons decreases, e.g. to \unit[50]{\%}
for $1/R=\unit[1.5]{TeV}$~\cite{Macesanu:2002ew}.

The UED model considered here is defined by specifying $R$ and
$\Lambda$, the ultraviolet cut-off used in the calculation of
radiative corrections to the KK masses. This analysis sets $\Lambda $
such that $\Lambda R = 20$. The $\gamma^*$ mass is insensitive to
$\Lambda$, while other KK masses typically change by a few per cent
when varying $\Lambda R$ in the range $10-30$. For $1/R =
\unit[1200]{GeV}$, the masses of the first-level KK photon, quark, and
gluon are $1200$, $1387$ and \unit[1468]{GeV},
respectively~\cite{ElKacimi:2009zj}.
Further details of the model are given in \Ref~\cite{Aad:2011kz}.
 
\section{Simulated samples}

For the GGM model, the SUSY mass spectra were calculated using
\Suspect 2.41~\cite{Djouadi:2002ze} and
\Sdecay~1.3~\cite{Muhlleitner:2003vg}.  The Monte Carlo~(MC) signal
samples were produced using \Pythia~6.423~\cite{pythia} with {\tt
  MRST2007 LO${}^{*}$}~\cite{Sherstnev:2007nd} parton distribution
functions~(PDF).
Cross sections were calculated at next-to-leading order (NLO) using
\Prospino 2.1~\cite{prospino2,Beenakker:1996ch}.
For the SPS8 model, the SUSY mass spectra were calculated using
\Isajet~7.80~\cite{isajet}. The MC signal samples were produced using
\Herwigpp~2.4.2~\cite{Bahr:2008pv} with {\tt MRST2007 LO${}^{*}$} PDF.
NLO cross sections were calculated using \Prospino.
In the case of the UED model, MC signal samples were generated using
the UED model as implemented at leading order (LO) in
\Pythia~\cite{ElKacimi:2009zj}.

The ``irreducible'' background from $(W\to\ell\nu)\gamma\gamma$ and
$(Z\rightarrow\nu\nu)\gamma\gamma$ production was simulated at LO
using \Madgraph~4~\cite{Alwall:2007st} with {\tt
  CTEQ6L1}~\cite{cteq6m} PDF. Parton showering and fragmentation were
simulated with \Pythia. NLO cross sections and scale uncertainties
from \Ref~\cite{Bozzi:2011en,Bozzi:2011wwa} were used.
In all cases the underlying event was simulated within the respective
generator.

All samples were processed through the {\tt GEANT4}-based
simulation~\cite{geant4} of the ATLAS detector~\cite{Aad:2010wq}. In
addition, the signal samples were overlaid with simulated minimum bias
events to model the average number of six $pp$ interactions per bunch
crossing (pile-up) experienced during the considered data-taking
period. More details may be found in \Ref~\cite{Aad:2011kz}.

\section{ATLAS detector}
The ATLAS detector~\cite{Aad:2008zzm} is a multi-purpose apparatus
with a forward-backward symmetric cylindrical geometry and nearly
4$\pi$ solid angle coverage.  Closest to the beamline are tracking
devices comprised of layers of silicon-based pixel and strip detectors
covering $\left|\eta\right|<2.5$\footnote{ATLAS uses a right-handed
  coordinate system with its origin at the nominal interaction point
  (IP) in the centre of the detector and the $z$-axis along the beam
  pipe. The $x$-axis points from the IP to the centre of the LHC ring,
  and the $y$-axis points upward. Cylindrical coordinates $(R,\phi)$
  are used in the transverse plane, $\phi$ being the azimuthal angle
  around the beam pipe. The pseudorapidity is defined in terms of the
  polar angle $\theta$ as $\eta=-\ln\tan(\theta/2)$.} and straw-tube
detectors covering $\left|\eta\right|<2.0$, located inside a thin
superconducting solenoid that provides a \unit[2]{T} magnetic field.
The straw-tube detectors also provide discrimination between electrons
and charged hadrons based on transition radiation.  Outside the
solenoid, fine-granularity lead/liquid-argon (LAr) electromagnetic
(EM) calorimeters provide coverage for $\left|\eta\right| < 3.2$ to
measure the energy and position of electrons and photons.  In the
region $\left|\eta\right| < 2.5$, the EM calorimeters are segmented
into three layers in depth.  The second layer, in which most of the EM
shower energy is deposited, is divided into cells of granularity of
$\Delta \eta \times \Delta \phi = 0.025 \times 0.025$. The first layer
is segmented with finer granularity to provide discrimination between
single photons and overlapping photons coming from the decays of
neutral mesons.  A presampler, covering $\left|\eta\right| < 1.8$, is
used to correct for energy lost upstream of the EM calorimeter.  An
iron/scintillating-tile hadronic calorimeter covers the region $|\eta|
< 1.7$, while copper and liquid-argon technology is used for hadronic
calorimeters in the end-cap region $1.5 < |\eta| < 3.2$. In the
forward region $3.2 < |\eta| < 4.5$ liquid-argon calorimeters with
copper and tungsten absorbers measure the electromagnetic and hadronic
energy.  A muon spectrometer consisting of three superconducting
toroidal magnet systems, tracking chambers, and detectors for
triggering surrounds the calorimeter system.

\section{Object reconstruction}
The reconstruction of converted and unconverted photons and of
electrons is described in Refs.~\cite{Aad:2010sp} and
\cite{Aad:2011mk}, respectively.

Converted photons have EM calorimeter clusters matched to tracks
coming from a conversion vertex. A conversion vertex is either a
vertex that has two tracks with large transition radiation in the
straw-tube detector and an invariant mass of the two tracks consistent
with a massless particle, i.e. a photon, or one track with large
transition radiation that has no associated hits in the pixel layer
closest to the beam line.  Electrons have a track matched to the EM
calorimeter cluster, and the track must have hits in the silicon
detectors, momentum not smaller than one tenth the cluster energy, and
transverse momentum of at least $\unit[2]{GeV}$.  Clusters matched to
neither a track or tracks coming from a conversion vertex nor an
electron track as described above are classified as unconverted
photons. A heuristic using the pixel hits closest to the beam line and
the track momenta is applied to choose between the photon and electron
interpretation in cases where the object can be both.

Photon candidates were required to be within $\left|\eta\right| <
1.81$, the value being chosen by an optimisation of the signal
acceptance versus background rejection, and to be outside the
transition region $1.37 < \left|\eta\right| < 1.52$ between the barrel
and the end-cap calorimeters.  The analysis used ``loose'' and
``tight'' photon selections~\cite{Aad:2010sp}.  The loose photon
selection includes a limit on the fraction of the energy deposit in
the hadronic calorimeter as well as a requirement that the transverse
width of the shower, measured in the middle layer of the EM
calorimeter, be consistent with the narrow shape expected for an EM
shower.  The tight photon selection additionally uses shape
information from the first layer to distinguish between isolated
photons and photons from the decay of neutral mesons.

The reconstruction of \MET is based on energy deposits in calorimeter
cells inside three-dimensional clusters with $|\eta | < 4.5$ and is
corrected for contributions from muons, if any~\cite{Aad:2011re}. The
cluster energy is calibrated to correct for the non-compensating
calorimeter response, energy loss in dead material, and out-of-cluster
energy.

Jets were reconstructed using the anti-$k_t$ jet
algorithm~\cite{Cacciari:2008gp} with four-momentum recombination and
radius parameter $R=0.4$ in $\eta$-$\phi$ space. They were required to
have $\pt > \unit[25]{GeV}$ and $|\eta| < 2.8$.

\section{Data analysis}

The data sample, corresponding to an integrated luminosity of
\integLumiE, was selected by a trigger requiring two loose photon
candidates with a transverse energy ($\ET$) above \unit[20]{GeV}.
In the offline analysis events were retained if they contained at
least two tight photon candidates with $\ET > \unit[25]{GeV}$.  In
addition, a photon isolation cut was applied, whereby the \ET deposit
in a cone of radius $0.2$ in the $\eta$-$\phi$ space around the centre
of the cluster, excluding the cells belonging to the cluster, had to
be less than \unit[5]{GeV}. The \ET was corrected for leakage from the
photon energy outside the cluster and for soft energy deposits from
pile-up interactions.
A cut of $\MET>\unit[125]{GeV}$~\cite{Aad:2011kz} defined the signal
region.  Preference was given to a common signal region for the three
models considered.

\begin{figure}[tb]
  \begin{center}
    \includegraphics[width=0.48\textwidth]{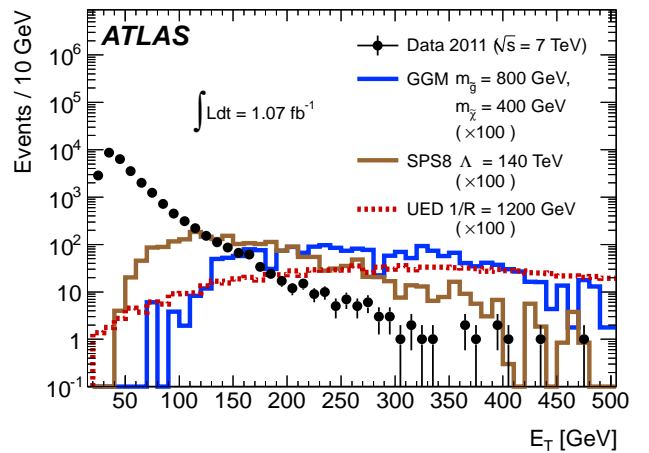}
  \end{center}
  \caption{The \ET spectrum of the leading photon in the $\gamma\gamma$
    candidate events in the data (points, statistical uncertainty only) together with the spectra from simulated GGM
    ($\mass{\gluino},\mass{\neutralino} = \unit[(800,400)]{GeV}$), SPS8
    ($\Lambda = \unit[140]{TeV}$), and UED
    ($1/R = \unit[1200]{GeV}$) samples, prior to the application of the 
    $\MET>\unit[125]{GeV}$ cut.  The signal samples are scaled by a factor of 100 for clarity.
    \label{fig:PhoPt}
  }
\end{figure}

A total of 27293 $\gamma\gamma$ candidate events were observed passing
all selections except the \MET cut.  The \ET distribution of the
leading photon for events in this sample is shown in
\Fig~\ref{fig:PhoPt}. Also shown are the \ET spectra obtained from GGM
MC samples for $\mass{\gluino}=\unit[800]{GeV}$ and
$\mass{\neutralino}=\unit[400]{GeV}$, from SPS8 MC samples with
$\Lambda = \unit[140]{TeV}$, and from UED MC samples for $1/R =
\unit[1200]{GeV}$, representing model parameters near the expected
exclusion limit. After the $\MET > \unit[125]{GeV}$ cut, \Nsignal
candidate events survived.

\begin{figure}[tb]
\begin{center}
\includegraphics[width=0.48\textwidth]{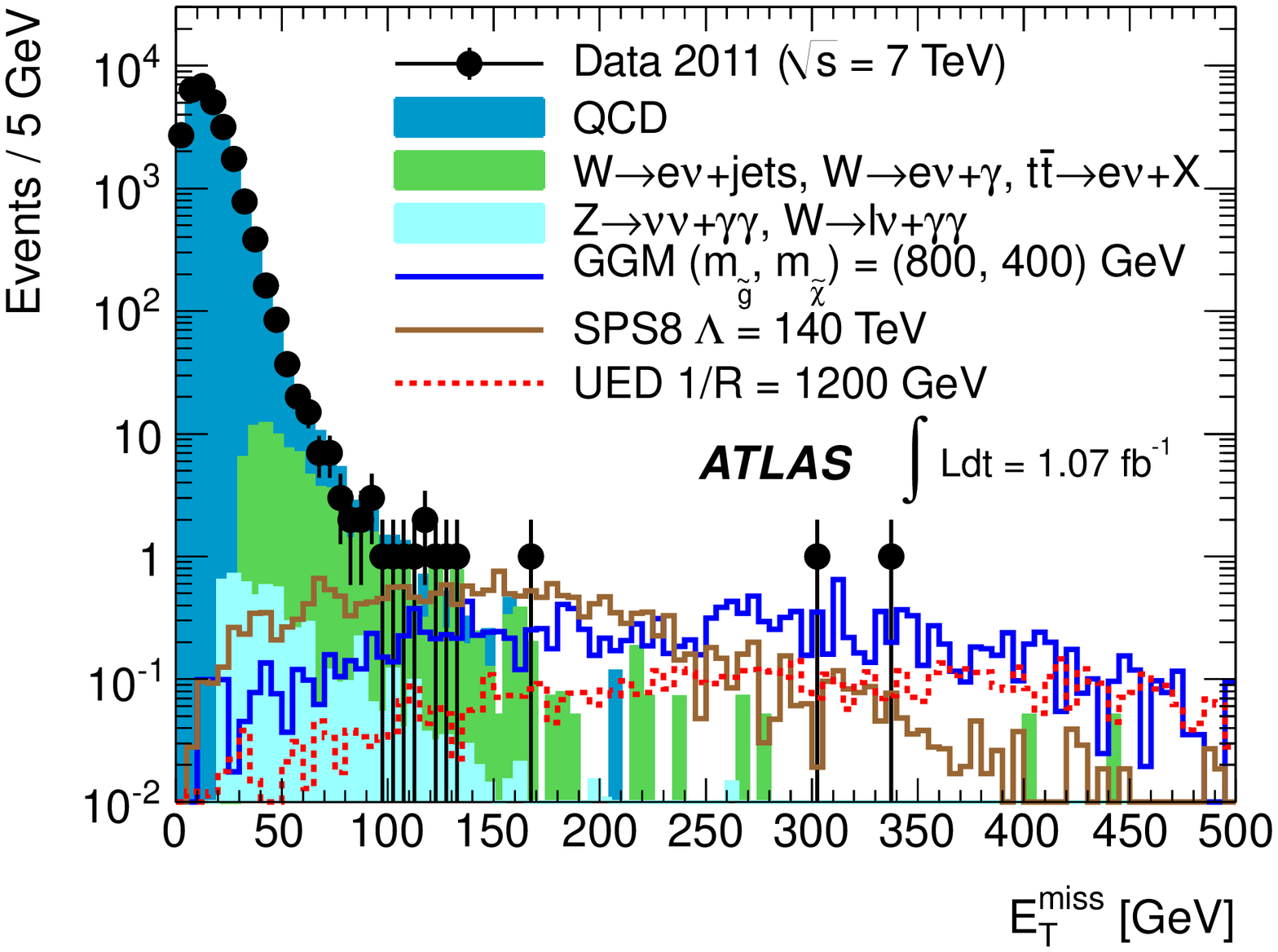}
\end{center}
\caption{\MET spectra for the $\gamma\gamma$ candidate events in data (points,
  statistical uncertainty only) and the estimated QCD background
  (normalised to the number of $\gamma\gamma$ candidates with $\MET <
  \unit[20]{GeV}$), the $\myW$ and $\myttbar$ backgrounds as estimated from
  the electron-photon control sample, and the irreducible
  background of \Zgg and \Wgg. Also shown are the expected signals
  from GGM ($\mass{\gluino},\mass{\neutralino} =
  \unit[(800,400)]{GeV}$), SPS8 ($\Lambda = \unit[140]{TeV}$), and UED
  ($1/R = \unit[1200]{GeV}$) samples.
  \label{fig:bkgnd}
}
\end{figure}

\section{Background estimation}
Following the procedure described in \Ref~\cite{Aad:2011kz}, the
contribution to large~\MET diphoton events from SM sources can be
grouped into two primary components and estimated with dedicated
control samples using data.  The first of these components, referred
to as ``QCD background'' for brevity, arises from a mixture of
processes that include $\gamma \gamma$ production as well as $\gamma $
+ jet and multijet events with at least one jet mis-reconstructed as a
photon.  The second background component is due to $W+X$ and \ttbar
events, where mis-reconstructed photons can arise from electrons and jets, for
which final-state neutrinos produce significant \met.

In order to estimate the QCD background from $\gamma\gamma$, $\gamma$
+ jet, and multijet events, a ``\QCDg control sample'' was extracted
from the diphoton trigger sample by selecting events for which at
least one of the photon candidates does not pass the tight photon
identification. Electrons were vetoed to remove contamination from
$W\to e\nu$ decays.  
%
The QCD background contamination in the signal region $\MET >
\unit[125]{GeV}$ was obtained from this QCD template after normalising
it to data in the region $\MET < \unit[20]{GeV}$. This gives a QCD
background expectation in the signal region of
$0.8\pm0.3(\mathrm{stat})$ events.
An alternate model for the QCD background was
obtained using a sample of dielectron events, with no jets, selected
by requiring two electrons with $\ET>\unit[25]{GeV}$ and
$\left|\eta\right| < 1.81$ and an invariant mass consistent with the
$Z$ boson mass.
As confirmed by MC simulation, the \MET spectrum of this $\Zee$ sample
with no additional jets, which is dominated by the calorimeter
response to two genuine EM objects, accurately represents the \MET
spectrum of SM $\gamma\gamma$ events. This spectrum was normalised in
the same way as the QCD control sample.
A systematic uncertainty of 0.6 events was assigned as the systematic
uncertainty on the background prediction from the relative fractions
of $\gamma\gamma$, $\gamma$ + jet, and multijet events using the
difference between the background estimates obtained using the QCD and
the \Zee templates, yielding the result of
$0.8\pm0.3(\mathrm{stat})\pm0.6(\mathrm{syst})$ events.
%
%
The \MET spectra of the QCD background and the $\gamma\gamma$ sample
are shown in \Fig~\ref{fig:bkgnd}.

\begin{figure}[bt]
\begin{center}
\includegraphics[width=0.48\textwidth]{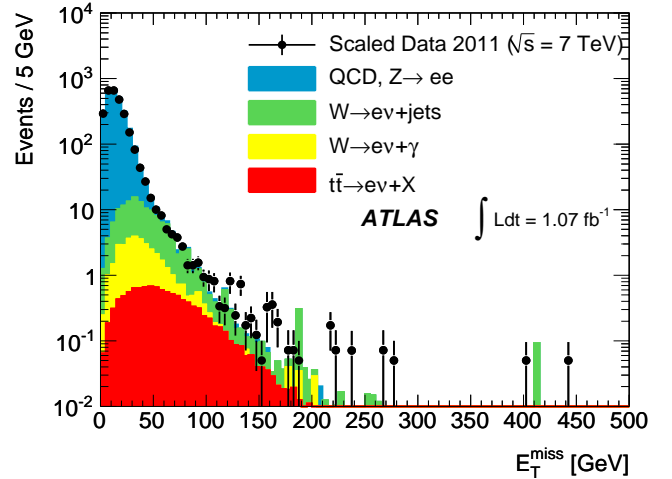}
\end{center}
\caption{\MET spectrum for the electron-photon control sample in data (points, statistical uncertainty only), normalised according to the probability for an  
electron to be mis-reconstructed as a tight photon, compared to the expected
backgrounds displayed by components (stacked histograms).
For the purpose of this comparison, the 
expected contributions from $\myW$ and \myttbar events are taken from 
MC simulation.}
\label{fig:ElPho}
\end{figure}

\begin{table*}[t!]
\caption{Number of observed $\gamma\gamma$ candidates in various \MET
  ranges in the data, as well as the expected numbers of SM background
  events estimated from the QCD and electron-photon control samples
  and, for the irreducible \Zgg and \Wgg processes, from MC
  simulation. Also shown are the expected numbers of signal events
  from GGM with $(\mass{\gluino},\mass{\neutralino}) =
  \unit[(800,400)]{GeV}$, SPS8  with $\Lambda = \unit[140]{TeV}$, and
  UED with $1/R =   \unit[1200]{GeV}$. The uncertainties are
  statistical only.  The $\MET < \unit[20]{GeV}$ region (first row) is
  used to normalise the QCD background to the   number of observed
  $\gamma \gamma$ candidates.
} 
\label{tab:events}
\center
\begin{tabular}{c|r|cccc|ccc} 
\hline
\hspace{-0.06in}\met range  & \multicolumn{1}{c|}{Data}  & \multicolumn{4}{c|}{Predicted background events} & \multicolumn{3}{c}{Expected signal events} \\
   $[$GeV$]$  & events & Total & QCD & \makebox[57pt]{$W/\ttbar(\rightarrow e \nu)+X$} & Irreducible  & GGM & SPS8 & UED \\ \hline

  \hspace{0.07in}0 - 20    & 20881 & \hspace{0.13in}-             & \hspace{0.01in}-            &  \hspace{0.06in}-            & -              & $\hspace{0.06in}0.20\pm 0.05$  & $\hspace{0.06in}0.22\pm 0.04$ & $0.02\pm 0.01$  \\ \hline
  \hspace{0.00in}20 - 50   &  6304 & $\hspace{0.00in}5968\pm 29$  & $\hspace{-0.12in}5951\pm28$  & $\hspace{0.00in}13.3\pm 8.1$ & $3.55\pm 0.35$ & $\hspace{0.06in}0.45 \pm 0.08$ & $\hspace{0.06in}1.53\pm 0.10$ & $0.11\pm 0.01$ \\
  \hspace{0.00in}50 - 75   &    86 & $\hspace{0.06in}87.1\pm 3.3$ & $\hspace{-0.06in}60.9\pm2.8$ & $\hspace{0.00in}25.2\pm 1.7$ & $1.01\pm 0.16$ & $\hspace{0.06in}0.48 \pm 0.08$ & $\hspace{0.06in}2.19\pm 0.12$ & $0.14\pm 0.01$ \\
  \hspace{0.06in}75 - 100  &    11 & $\hspace{0.06in}14.7\pm 1.2$ & $\hspace{0.0in}6.7\pm0.9$  & $\hspace{0.06in}7.4\pm 0.8$  & $0.52\pm 0.10$ & $\hspace{0.06in}0.75 \pm 0.10$ & $\hspace{0.06in}2.09\pm 0.11$ & $0.15\pm 0.01$ \\
  \hspace{0.00in}100 - 125 &     6 & $\hspace{0.12in}4.9\pm 0.7$  & $\hspace{0.0in}1.6\pm0.4$  & $\hspace{0.06in}3.0\pm 0.5$  & $0.32\pm 0.08$ & $\hspace{0.06in}1.20 \pm 0.12$ & $\hspace{0.06in}2.53\pm 0.13$ & $0.29\pm 0.02$ \\
  \hspace{0.22in}$>$ 125   &     5 & $\hspace{0.12in}4.1\pm 0.6$  & $\hspace{0.0in}0.8\pm 0.3$ & $\hspace{0.06in}3.1\pm 0.5$  & $0.23\pm 0.05$ &                $17.2 \pm 0.5$  &               $12.98\pm 0.28$  & $9.67\pm 0.11$ \\
\hline
\end{tabular}
\end{table*}

The second significant background contribution, from $W+X$ and \ttbar
events, was estimated via an ``electron-photon'' control sample
composed of events with at least one photon and one electron, each with
$\ET>\unit[25]{GeV}$, and scaled by the probability for an electron to
be mis-reconstructed as a tight photon, as estimated from a study of the
$Z$ boson in the $ee$ and $e\gamma$ sample. The scaling factor varies
between \unit[5]{\%} and \unit[17]{\%} as a function of $\eta$, since
it depends on the amount of material in front of the calorimeter.
Events with two or more photons were vetoed from the control sample to keep it
orthogonal to the signal sample. In case of more than one electron,
the one with the highest \pt was used.
The \MET spectrum for the scaled electron-photon control sample is
shown in \Fig~\ref{fig:ElPho}, where it is compared to the expected
contributions from various background sources as computed from MC
simulation.  The electron-photon control sample has a significant
contamination from \Zee events, in which one electron is mis-reconstructed as a photon,
and from QCD processes mentioned above. Both of these contaminations
must be subtracted in order to extract the contribution to the \MET
distribution from events with genuine \MET, such as $W+X$ and \ttbar.
The contribution from QCD and \Zee events was estimated by normalising
the \QCDg control sample to the scaled electron-photon \MET
distribution in the region $\MET < \unit[20]{GeV}$ where they
dominate, as shown in \Fig~\ref{fig:ElPho}.  This distribution was
then subtracted from the scaled electron-photon control sample,
yielding a prediction for the contribution to the high-\MET diphoton
sample from $W+X$ and \ttbar events.  This procedure led to an
estimate of the background from $W+X$ and \ttbar production of $3.1
\pm 0.5(\mathrm{stat})$ events in the signal region. A systematic
uncertainty of 0.06 events was assigned by using the \Zee template in
place of the \QCDg template when subtracting the contamination due to
\Zee and QCD processes.
The contribution from $WW$ events to the electron-photon control
sample was estimated using MC simulation and found to be negligible.

A parallel study using MC samples of \myW and \myttbar, rather than
the electron-photon control sample, gave an estimate of $1.8 \pm
1.2(\mathrm{stat})$ background events.  The difference was taken as an
estimate of the systematic uncertainty, yielding the result of
$3.1\pm0.5(\mathrm{stat})\pm1.4(\mathrm{syst})$ events.  Also included
in the quoted systematic uncertainty is the relative uncertainty ($\pm
\unit[10]{\%}$) on the probability for an electron to be mis-reconstructed as a photon.

A small irreducible background of $0.23 \pm 0.05 (\mathrm{stat}) \pm
0.04 (\mathrm{syst})$ events from \Zgg and \Wgg events was estimated
from MC simulation. The systematic uncertainty accounts for variations
in the factorisation and renormalisation scales in the NLO
calculations.
The contamination from cosmic-ray muons was found to be negligible.

Figure~\ref{fig:bkgnd} shows the \MET spectrum of the selected $\gamma
\gamma$ candidates, superimposed on the estimated backgrounds.
Table~\ref{tab:events} summarises the number of observed $\gamma
\gamma$ candidates, the expected backgrounds, and three representative
GGM, SPS8, and UED signal expectations, in several \MET ranges.  No
indication of an excess at high \MET values, where the signal is
expected to dominate, is observed.

\section{Signal efficiencies and systematic uncertainties}
The GGM signal efficiency was determined using MC simulation over an
area of the GGM parameter space that ranges from \unit[400]{GeV} to
\unit[1200]{GeV} for the gluino mass, and from $\unit[50]{GeV}$ to
within $\unit[20]{GeV}$ of the gluino mass for the neutralino mass.
The efficiency increases smoothly from $\unit[5.5]{\%}$ to
$\unit[31]{\%}$ for $(\mass{\gluino},\mass{\neutralino}) =
\unit[(400,50)]{GeV}$ to $\unit[(1200,1100)]{GeV}$.
The SPS8 signal efficiency increases smoothly from \unit[9.2]{\%}
($\Lambda = \unit[80]{TeV}$) to \unit[29.4]{\%} ($\Lambda =
\unit[220]{TeV}$).
The UED signal efficiency, also determined using MC simulation,
increases smoothly from \unit[48.9]{\%} ($1/R = \unit[1000]{GeV}$) to
\unit[52.6]{\%} ($1/R = \unit[1500]{GeV}$).

\begin{table}[b!]
  \caption{Relative systematic uncertainties on the expected signal
    yield for GGM with $(\mass{\gluino},\mass{\neutralino}) = 
    \unit[(800,400)]{GeV}$, SPS8 with $\Lambda=\unit[140]{TeV}$, and UED with $1/R = \unit[1200]{GeV}$. No
    PDF and scale uncertainties are given for the UED case as the cross
    section is evaluated to LO. }
  \label{tab:systematics}
  \center
\begin{tabular}{lrrr} 
  \hline
  Source of uncertainty & \multicolumn{3}{c}{Uncertainty} \\
& \multicolumn{1}{c}{GGM} & \multicolumn{1}{c}{SPS8} & \multicolumn{1}{c}{UED} \\ \hline
 Integrated luminosity           & $\hspace{0.04in}3.7$\% & $\hspace{0.04in}3.7$\% & $\hspace{0.04in}3.7$\% \\
 Trigger                         & $\hspace{0.04in}0.6$\% & $\hspace{0.04in}0.6$\% & $\hspace{0.04in}0.6$\% \\
 Photon identification           & $\hspace{0.02in}3.9$\% & $\hspace{0.02in}3.9$\% & $\hspace{0.02in}3.7$\% \\
 Photon isolation                & $\hspace{0.02in}0.6$\% & $\hspace{0.02in}0.6$\% & $\hspace{0.02in}0.5$\% \\
 Pile-up                          & $\hspace{0.02in}1.3$\% & $\hspace{0.02in}1.3$\% & $\hspace{0.02in}1.6$\% \\
 $\MET$ reconstruction and scale & $\hspace{0.02in}1.7$\% & $\hspace{0.02in}5.6$\% & $\hspace{0.02in}0.7$\% \\
 LAr readout                     & $\hspace{0.02in}1.0$\% & $\hspace{0.02in}0.7$\% & $\hspace{0.02in}0.4$\% \\
 Signal MC statistics            & $\hspace{0.02in}2.9$\% & $\hspace{0.02in}2.3$\% & $\hspace{0.02in}1.8$\% \\
 \hline
 Total signal uncertainty        & $\hspace{0.02in}6.6$\% & $\hspace{0.02in}8.3$\%&  $\hspace{0.02in}6.0$\% \\
 \hline
 PDF and scale                   & $\hspace{0.02in}31$\% & $\hspace{0.02in}5.5$\% & $\hspace{0.02in}-$  \\ 
 \hline
 Total                           & $\hspace{0.02in}32$\% & $\hspace{0.02in}10$\% & $\hspace{0.02in}6.0$\% \\
 \hline
 \end{tabular}
\end{table}

The various relative systematic uncertainties on the GGM, SPS8, and UED
signal cross sections are summarised in \Tab~\ref{tab:systematics} for
the chosen GGM, SPS8, and UED reference points. The uncertainty on the
luminosity is \unit[3.7]{\%}~\cite{lumi2011,Aad:2011dr}.
The trigger efficiency of the required diphoton trigger was estimated from
the efficiency of the corresponding single photon trigger, which was
estimated using a bootstrap method~\cite{Aad:2011xs}. The result
is $99.92^{+0.04}_{-0.18} \%$ for events passing all selections except
the final \MET cut.
To estimate the systematic uncertainty due to the unknown composition
of the data sample, the trigger efficiency was also evaluated on MC
events using mis-reconstructed photons from filtered multijet samples and photons
from signal (SUSY and UED) samples. A conservative systematic
uncertainty of \unit[0.6]{\%} was derived from the difference between
the obtained efficiencies. Uncertainties on the photon selection, the
photon energy scale, and the detailed material composition of the
detector, as described in \Ref~\cite{Aad:2011kz}, result in an
uncertainty of \unit[3.9]{\%} for the GGM and SPS8 signals and
\unit[3.7]{\%} for the UED signal. The uncertainty from the photon
isolation was estimated by varying the energy leakage and the pile-up
corrections independently, resulting in an uncertainty of
\unit[0.6]{\%} for GGM and SPS8 and \unit[0.5]{\%} for UED.  The
influence of pile-up on the signal efficiency, evaluated by comparing
GGM/SPS8 (UED) MC samples with different pile-up configurations, leads
to a systematic uncertainty of \unit[1.3]{\%}(\unit[1.6]{\%}).
Systematic uncertainties due to the $\MET$ reconstruction, estimated
by varying the cluster energies within established ranges and the
$\MET$ resolution between the measured performance and MC
expectations, contribute an uncertainty of \unit[0.1]{\%} to
\unit[12.4]{\%}~(GGM), \unit[1.7]{\%} to \unit[13.8]{\%} (SPS8), and
\unit[0.5]{\%} to \unit[1.5]{\%}~(UED). A systematic uncertainty was
also assigned to account for temporary failures of the LAr calorimeter
readout during part of the data-taking period, which was not modelled
in the MC samples. Electrons and photons were removed from the
afflicted area, but jets, being larger objects, were not. Jet energy
corrections were therefore applied. Varying these corrections over
their range of uncertainty results in systematic uncertainties of
\unit[1.0]{\%}, \unit[0.7]{\%}, and \unit[0.4]{\%} for GGM, SPS8, and
UED, respectively.
Added in quadrature, the total systematic uncertainty on the signal
yield varies between \unit[6.3]{\%} and \unit[15]{\%}~(GGM),
\unit[6.2]{\%} and \unit[15]{\%}~(SPS8), and \unit[5.8]{\%} and
\unit[6.0]{\%} (UED).

The PDF uncertainties on the GGM (SPS8) cross sections were evaluated
by using the {\tt CTEQ6.6M} PDF error sets~\cite{Stump:2003yu} in the
\Prospino cross section calculation and range from \unit[12]{\%} to
\unit[44]{\%} (\unit[4.7]{\%} to \unit[6.6]{\%}).  The factorisation
and renormalisation scales in the NLO \Prospino calculation were
increased and decreased by a factor of two, leading to a systematic
uncertainty between \unit[16]{\%} and \unit[23]{\%} (\unit[1.7]{\%}
and \unit[6.7]{\%}) on the expected cross sections.
The different impact of the PDF and scale uncertainties of the GGM and
SPS8 yields is related to the different production mechanisms in the
two models (see \Section~\ref{sec:susy}).
In the case of UED, the PDF uncertainties were evaluated by using the
{\tt MSTW2008 LO}~\cite{MSTW2008} PDF error sets in the LO cross
section calculation and are about \unit[4]{\%}. The scale of
$\alpha_s$ in the LO cross section calculation was increased and
decreased by a factor of two, leading to a systematic uncertainty of
\unit[4.5]{\%} and \unit[9]{\%}, respectively. NLO calculations are
not yet available, but are expected to be much larger than the PDF and
scale uncertainties. Thus, the LO cross sections were used for the
limit calculation without any theoretical uncertainty, and the effect
of PDF and scale uncertainties on the final limit is given separately.

\begin{figure}[t]
  \begin{center}
    \includegraphics[width=0.48\textwidth]{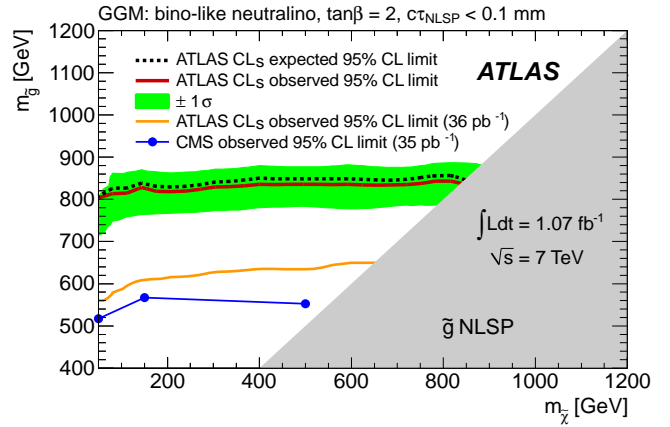}
  \end{center}
  \caption{Expected and observed \unit[95]{\%} \CL lower limits on the
    gluino mass as a function of the neutralino mass in the GGM model
    with a bino-like lightest neutralino NLSP (the grey area
    indicates the region where the NLSP is the gluino, which is not
    considered here). The other sparticle masses are fixed to
    $\sim\unit[1.5]{TeV}$. Further model parameters are $\tan\beta=2$
    and $c\tau_{\mathrm{NLSP}} < \unit[0.1]{mm}$. The
    previous ATLAS~\protect\cite{Aad:2011kz} and
    CMS~\protect\cite{Chatrchyan:2011wc} limits are also shown.
    \label{fig:ggm:limit}
  }
\end{figure}

\begin{figure}[t]
  \begin{center}
    \includegraphics[width=0.44\textwidth]{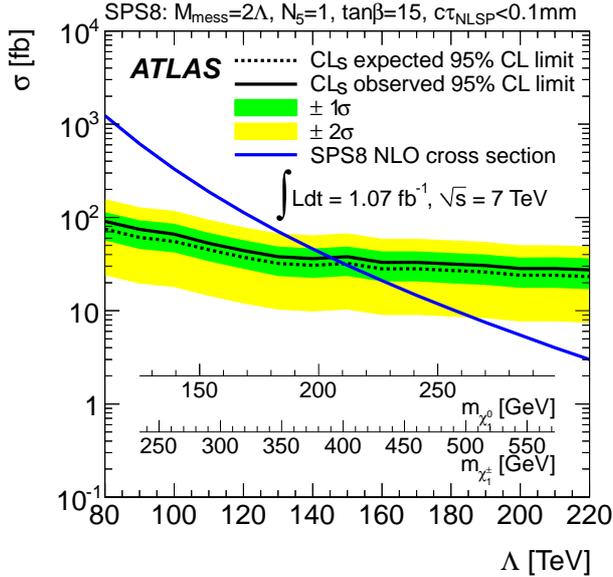}
  \end{center}
  \caption{Expected and observed \unit[95]{\%}~\CL upper limits on the
    sparticle production cross section in the SPS8 model, and the NLO cross section
    prediction, as a function of $\Lambda$ and the lightest neutralino
    and chargino masses.  Further SPS8 model parameters
    are $M_\mathrm{mess}=2\Lambda$, $\Nfive=1$, $\tan\beta=15$, and
      $c\tau_{\mathrm{NLSP}} < \unit[0.1]{mm}$.
      \label{fig:sps8:limit}
    }
\end{figure}

\begin{figure}[t]
  \begin{center}
    \includegraphics[width=0.44\textwidth]{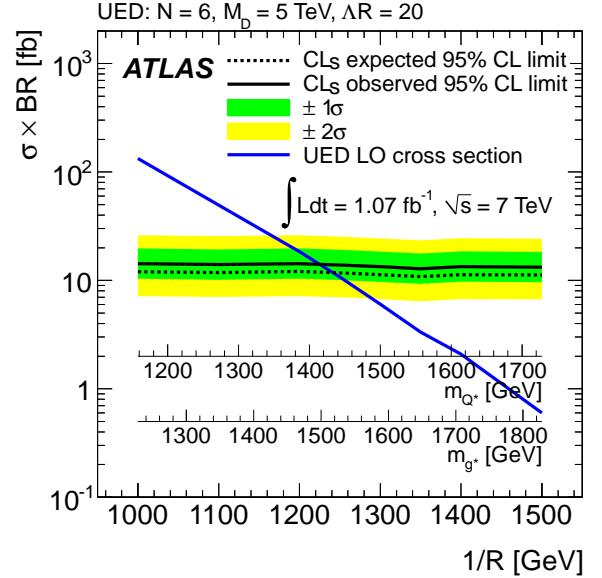}
  \end{center}
  \caption{Expected and observed \unit[95]{\%}~\CL upper limits on the
    KK particle production cross section times branching fraction to
    two photons in the UED model, and the LO
    cross section prediction times branching fraction,
    as a function of $1/R$ and the KK quark ($Q^{*}$) and KK gluon ($g^{*}$) masses.  The UED model parameters are $N=6$, $M_D
    = \unit[5]{TeV}$, and $\Lambda R = 20$. }
  \label{fig:ued:limit}
\end{figure}

\section{Results}
Based on the observation of \Nsignal events with
$\MET>\unit[125]{GeV}$ and a background expectation of \nbackground
events, a \unit[95]{\%}~\CL upper limit is set on the number of
events in the signal region from any scenario of physics beyond the SM 
using the profile likelihood and $CL_s$ method~\cite{Read:2002hq}.
The result is \Nlimit events at \unit[95]{\%}~\CL.

Further, \unit[95]{\%}~\CL upper limits on the cross sections of the
considered models are calculated, including all systematic
uncertainties except for theory uncertainties, i.e. PDF and scale. In
the GGM model the upper limit on the cross section is
\ensuremath{\unit[(22-129)]{fb}}\xspace, where the larger value
corresponds to $\mass{\gluino},\mass{\neutralino} =
\unit[(400,50)]{GeV}$. For $\mass{\neutralino}\ge\unit[150]{GeV}$, the
limit is below \unit[30]{fb}, reaching \unit[22]{fb} for heavy
neutralino masses. Figure~\ref{fig:ggm:limit} shows the expected and
observed lower limits on the GGM gluino mass as a function of the
neutralino mass. For comparison the lower limits from
ATLAS~\cite{Aad:2011kz} and CMS~\cite{Chatrchyan:2011wc} based on the
2010 data are also shown. The total systematic uncertainty includes
the theory uncertainties, which are dominant. Excluding the
PDF and scale uncertainty in the limit calculation would improve the
observed limit on the gluino mass by \unit[$\sim$10]{GeV}.

In the SPS8 model the cross section limit is \sigmaSPS as shown in
\Fig~\ref{fig:sps8:limit}, corresponding to
$\Lambda=\unit[220-80]{TeV}$. For illustration the cross section
dependence as a function of the lightest neutralino and chargino
masses is also shown. A lower limit on the SPS8 breaking scale
$\Lambda > \SPSlimit$ at \unit[95]{\%}~\CL is set including the theory
uncertainties, i.e. PDF and scale uncertainties, in the total
systematic uncertainty.

For the UED model the cross section limit is \sigmaUED for
$1/R=1000-\unit[1500]{GeV}$.  Figure~\ref{fig:ued:limit} shows the
limit on the cross section times branching ratio for the UED model,
which is \sigmaBRUED. For illustration the cross section dependence as
a function of the KK quark and KK gluon masses is also shown. A lower
limit on the UED compactification scale $1/R > \UEDlimitlow$ at
\unit[95]{\%}~\CL is set.
In this case PDF and scale uncertainties are not included when
calculating the limits. Including PDF and scale uncertainties computed
at LO degrade the limit on $1/R$ by a few \unit{GeV}.

\section{Conclusions}
A search for events with two photons and $\MET>\unit[125]{GeV}$,
performed using \integLumi of \unit[7]{TeV} $pp$ collision data
recorded with the ATLAS detector at the LHC, found \Nsignal events
with an expected background of \nbackground.
The results are used to set a model-independent \unit[95]{\%}~\CL
upper limit of \Nlimit events from new physics.  Upper limits at
\unit[95]{\%}~\CL are also set on the production cross section for
three particular models of new physics: \sigmaGGM for the GGM model,
\sigmaSPS for the SPS8 model, and \sigmaUED for the UED model.
Under the GGM hypothesis, a lower limit on the gluino mass of
\GGMlimit is determined for bino masses above \unit[50]{GeV}.  A lower
limit of \SPSlimit is set on the SPS8 breaking scale $\Lambda$, which
is the first limit on the SPS8 model at the LHC.  A lower limit of
\UEDlimitlow is set on the UED compactification scale $1/R$. These
results provide the most stringent tests of these models to date,
significantly improving upon previous best limits of
\unit[560]{GeV}~\cite{Aad:2011kz} for the GGM gluino mass,
\unit[124]{TeV}~\cite{Abazov:2010us} for $\Lambda$ in SPS8, and
\unit[961]{GeV}~\cite{Aad:2011kz} for $1/R$ in UED, respectively.

\section*{Acknowledgements}
\input{Acknowledgement-09Mar11-collisions.tex}

\bibliographystyle{atlasnote}
\bibliography{DiPhotonMetPRL}

\clearpage
\onecolumn
\input{atlas_authlist}

\end{document}

%% file: Acknowledgement-09Mar11-collisions.tex
We thank CERN for the very successful operation of the LHC, as well as the
support staff from our institutions without whom ATLAS could not be
operated efficiently.

We acknowledge the support of ANPCyT, Argentina; YerPhI, Armenia; ARC,
Australia; BMWF, Austria; ANAS, Azerbaijan; SSTC, Belarus; CNPq and FAPESP,
Brazil; NSERC, NRC and CFI, Canada; CERN; CONICYT, Chile; CAS, MOST and
NSFC, China; COLCIENCIAS, Colombia; MSMT CR, MPO CR and VSC CR, Czech
Republic; DNRF, DNSRC and Lundbeck Foundation, Denmark; ARTEMIS, European
Union; IN2P3-CNRS, CEA-DSM/IRFU, France; GNAS, Georgia; \linebreak BMBF, DFG, HGF, MPG
and AvH Foundation, Germany; GSRT, Greece; ISF, MINERVA, GIF, DIP and
Benoziyo Center, Israel; INFN, Italy; MEXT and JSPS, Japan; CNRST, Morocco;
FOM and NWO, Netherlands; RCN, Norway; MNiSW, Poland; GRICES and FCT,
Portugal; MERYS (MECTS), Romania; MES of Russia and ROSATOM, Russian
Federation; JINR; MSTD, Serbia; MSSR, Slovakia; ARRS and MVZT, Slovenia;
DST/NRF, South Africa; MICINN, Spain; SRC and Wallenberg Foundation,
Sweden; SER, SNSF and Cantons of Bern and Geneva, Switzerland; NSC, Taiwan;
TAEK, Turkey; STFC, the Royal Society and Leverhulme Trust, United Kingdom;
DOE and NSF, United States of America.

The crucial computing support from all WLCG partners is acknowledged
gratefully, in particular from CERN and the ATLAS Tier-1 facilities at
TRIUMF (Canada), NDGF (Denmark, Norway, Sweden), CC-IN2P3 (France),
KIT/GridKA (Germany), INFN-CNAF (Italy), NL-T1 (Netherlands), PIC (Spain),
ASGC (Taiwan), RAL (UK) and BNL (USA) and in the Tier-2 facilities
worldwide.

%% file: atlas_authlist.tex
\begin{flushleft}
{\Large The ATLAS Collaboration}

\bigskip

G.~Aad$^{\rm 48}$,
B.~Abbott$^{\rm 111}$,
J.~Abdallah$^{\rm 11}$,
A.A.~Abdelalim$^{\rm 49}$,
A.~Abdesselam$^{\rm 118}$,
O.~Abdinov$^{\rm 10}$,
B.~Abi$^{\rm 112}$,
M.~Abolins$^{\rm 88}$,
H.~Abramowicz$^{\rm 153}$,
H.~Abreu$^{\rm 115}$
% ...
\bigskip

% ...
$^{10}$ Institute of Physics, Azerbaijan Academy of Sciences, Baku, Azerbaijan\\
$^{11}$ Institut de F\'isica d'Altes Energies and Departament de F\'isica de la Universitat Aut\`onoma  de Barcelona and ICREA, Barcelona, Spain\\
% ...
$^{48}$ Fakult\"{a}t f\"{u}r Mathematik und Physik, Albert-Ludwigs-Universit\"{a}t, Freiburg i.Br., Germany\\
$^{49}$ Section de Physique, Universit\'e de Gen\`eve, Geneva, Switzerland\\
% ...
$^{88}$ Department of Physics and Astronomy, Michigan State University, East Lansing MI, United States of America\\
% ...
$^{111}$ Homer L. Dodge Department of Physics and Astronomy, University of Oklahoma, Norman OK, United States of America\\
$^{112}$ Department of Physics, Oklahoma State University, Stillwater OK, United States of America\\
% ...
$^{115}$ LAL, Univ. Paris-Sud and CNRS/IN2P3, Orsay, France\\
% ...
$^{118}$ Department of Physics, Oxford University, Oxford, United Kingdom\\
% ...
$^{153}$ Raymond and Beverly Sackler School of Physics and Astronomy, Tel Aviv University, Tel Aviv, Israel\\
% ...
\end{flushleft}